# Targeted Code Inspection based on Human Errors

Fuqun Huang
*Centre for Informatics and Systems*
*University of Coimbra*
Coimbra, Portugal
huangfuqun@dei.uc.pt

Henrique Madeira
*Centre for Informatics and Systems*
*University of Coimbra*
Coimbra, Portugal
henrique@dei.uc.pt

*Abstract*— **As a direct cause of software defects, human error is the key to understanding and identifying defects. We propose a new code inspection method: targeted code inspection based on human error mechanisms of software engineers. Based on the common erroneous mechanisms of human cognition, the method targets error-prone codes with high efficiency and minimum effort. The proposed method is supported by preliminary evidence in a pilot study.**

## I. INTRODUCTION

As the primary cause of software defects, human error is the key to understanding and identifying defects. The human error nature of defects has not been explored until recently, by only a few pioneers in the world. To the best of our knowledge, there are four groups of researchers pioneering this field. Huang and Liu [1] first proposed the area "Software Fault Defense based on Human Errors" to systematically reduce software defects based on a deep understanding of the human errors of software practitioners. Huang et al. has carried out in-depth studies on a variety of topics, such as defect prevention based on human errors [2, 3], software diversity design based on human errors [4], and defect prediction based on human errors [5, 6]. Anu and Walia et al. have been focusing on using human error taxonomy to improve requirement quality [7]. Nagaria and Hall [8] interviewed developers about the situations of the skill-based errors and how they mitigate such errors. Couceiro and Madeira et al. [9] developed techniques to identify suspicious code by monitoring programmers' biometrics. These studies show that studying human errors of software engineers is beneficial. However, there lacks a human-error based method for code inspection, despite the fact that human errors are a direct cause of defects in code, while code review has been a significant part of the daily work for software engineers.

This paper proposes a **H**uman **E**rror-based **C**ode Ins**p**ection (HECIP) method, which targets at the error-prone code snippets with high efficiency and minimum effort. HECIP can be used as a good complementary to the established code review methods.

## II. CONCEPTS

**Defect:** an incorrect or missing step, process, or data definition in a computer program.

**Error:** an erroneous human behavior that leads to a software defect. Errors are classified at a finer-grained level in psychology as mistakes, slips or lapses [10].

**Error Mode:** a particular pattern of erroneous behavior that recurs across different activities, due to the cognitive mechanisms shared by all humans [10].

**Error-Prone Scenario (EPS)**: A set of conditions under which an Error Mode tends to occur. An EPS in code inspection mainly include the conditions around a programming task.

## III. THE HUMAN ERROR-BASED CODE INSPECTION METHOD

The HECIP method includes three components: Human Error Modes (HEM), Error-Prone Scenario, and Error-Prone Scenario Identification (EPSI).

### A. Error Modes

Psychologists find that human errors **take** a limited number of patterns across diverse contexts [10]. An example human error mode is described in the second row of Table I.

TABLE I. EXAMPLES OF HUMAN ERROR MODES AND ERROR-PRONE SCENARIOS

| **Error mode:** Difficulties with exponential developments [10] | | |
|---|---|---|
| "Processes that develop exponentially have great significance for systems in either growth or decline, yet subjects appeared to have no intuitive feeling for them. When asked to gauge such processes, they almost invariably underestimated their rate of change and were constantly surprised at their outcomes." This means humans tend to construct linear models (whose growth rate is lower than exponential models) when exponential models are required to understand a situation in reality [10]. | | |
| **EPS** | **IF** | *Current task requires* extracting a relation between independent variable x and dependent variable y according to a sample data; |
| | **WHEN** | The actual relation belongs to models in the families of "$y = x^p$" or "$y = d^x$"; |
| | **THEN** | People tend to construct wrong models in the Family of "$y = ax$". |
| **Error Mode**: Post-completion error [11] | | |
| If the ultimate goal is decomposed into sub-goals, a sub-goal is likely to be omitted under the following conditions: the sub-goal is not a necessary condition for achieving its super-ordinate goal, and the sub-goal is to be carried out at the end of the task. | | |
| **EPS** | **IF** | Task A ={Task A.1, Task A.2} |
| | **WHEN** | <Task A.1 is the main subtask>, AND <Task A.2 is not a necessary condition to Task A.1>, AND <Task A.2 is the last step of Task A >; |
| | **THEN** | Humans tend to omit Task A.2. |

Due to the page limit, Table I is *not* meant to exhaustively enumerate all of the human error modes described in cognitive psychology.

### B. Error-Prone Scenario

The fundamental psychological theories tend to be somewhat vague and thus difficult to apply for practical purposes. We extracted the Error-Prone Scenarios and represented as by pseudo codes, with examples shown in the third row of Table I.

*C. Error-Prone Scenario Identification*

EPSI is the activity of an inspector identifying the conditions in software development contexts and matching them to Error-Prone Scenarios. This process can be aided by a graphic tool that is adapted from Causal Mechanism Graph developed by Huang and Smidts [12], which is a powerful notation system to represent causal mechanisms.

## IV. PRELIMINARY EVIDENCE

The proposed idea is supported by preliminary evidence obtained in a pilot experiment.

*A. Process*

1). Two participants independent of this study volunteered as the HECIP code inspector. Participant #1 (P1) was a senior software developer who had over 10 years of industrial experience, and code review had been a routine in his daily work. Participant #2 (P2) was a young researcher who had just obtained her Ph.D. in computer Science and had no industrial experience.

2). The first author of this paper provided a short session of training lasting for 20 minutes on the HECIP method, mainly focused on explaining the two error modes and error-prone scenarios in Table I.

3). A program and its requirement (the "jiong" task) previously collected in [4] were provided to the participants of this study. The participants performed code inspection. The rule of the code inspection was to find all the defects as soon as possible in an un-interrupted time period. The participants recorded the start time of the code inspection, and the time when each defect was found.

*B. Results and discussions*

The results are shown in Table II.

TABLE II. RESULTS OF THE HECIP EFFICIENCY ASSESSMENT

| Participant | Time of finding Defect A[a] (minutes) | Time of identifying Defect B[b] (minutes) | Average Time of identifying other Defects (minutes) |
|---|---|---|---|
| #1 | 3 | 3 | 24.5 |
| #2 | 1 | 1 | 8.75 |

[a] Defect A: the relationship between the height and nest level is modeled wrongly as $h=8n$, instead of $h=2^{n+2}$. Defect A is a manifestion of the human error mode "Difficulties with exponential developments".
[b] Defect B: the blank line after a "jiong" is missing. Defect B is a manifestion of "Post-completion error".

Both participants identified the defects that manifest the Human Error Modes in the HECIP training--Defect A and B, in much less time than other defects. Notably, the participants found Defect A and B at the beginning of the code inspection process: P1 found A firstly and B secondly, out of a total of four defects; P2 found A secondly and B thirdly, out of six defects.

Interestingly, the more experienced participant P1 spent more time while found less number of defects than P2. The explanation is that P1 seemed to have fully understood the requirement and code and thought more deeply than P2: P1 devoted about half an hour in finding the third defect, which is a defect that appeared at five different locations in the code; whereas all of the defects P2 identified except for defect A and B were syntax defects. The program produced correct outputs after fixing the defects found by P1; conversely, it still failed after fixing the defects found by P2.

Nevertheless, both P1 and P2 identified Defect A (a difficult design defect) and Defect B (a defect originated from requirement) in impressive short periods than that of other defects. HECIP seems effective in helping the code inspectors identifying relevant defects with high efficiency.

This pilot study was intended to test the feasibility of HECIP, rather than provide a comprehensive evaluation on its benefit. A reader may notice that there was 20 minutes' training time, which could largely offset the time the inspectors saved in this small case. In potential applications, hours of HECIP training could help a software engineer identifying many defects and saving uncoutable time during one's professional life that is measured by tens of years.

## V. CONCLUSION

This paper proposed a new code inspection method based on human error theories. Preliminary evidence shows that the method helps code inspectors identifying targeted defects with high efficiency. A more comprehensive study involving more participants and more human error modes will follow up.